\documentclass[aps,prb,showpacs,floatfix,twocolumn,superscriptaddress,longbibliography]{revtex4-2}

\usepackage{float}
\usepackage{color}
\usepackage{bm}
\usepackage{todonotes}
\usepackage{verbatim}
\usepackage{soul}
\usepackage{glossaries}
\usepackage{sidecap}
\usepackage{hyperref} 
\hypersetup{
    colorlinks=true,
    linkcolor=blue,
    filecolor=magenta,      
    urlcolor=red,
    citecolor=blue,
}

\def\Q{\ensuremath{\mathbf{Q}}}
\def\TN{\ensuremath{T_\mathrm{N}}}
\def\Tc{\ensuremath{T_\mathrm{c}}}

\def\NN{nearest neighbor}
\def\FeSC{iron-based superconductors}

\newacronym{FWHM}{FWHM}{full-width at half-maximum}
\newacronym{AFM}{AFM}{antiferromagnetic}
\newacronym{FM}{FM}{ferromagnetic}
\newacronym{DM}{DM}{Dzyaloshinskii-Moriya}

\begin{document}
\title{Frustrated Magnetic Interactions and Quenched Spin Fluctuations in {CrAs}}

\author{Yayuan Qin}
\author{Yao Shen}\email[]{yshen.physics@gmail.com}
\author{Yiqing Hao}
\affiliation{State Key Laboratory of Surface Physics and Department of Physics, Fudan University, Shanghai 200433, China}

\author{Hongliang Wo}
\affiliation{State Key Laboratory of Surface Physics and Department of Physics, Fudan University, Shanghai 200433, China}
\affiliation{Shanghai Qizhi Institute, Shanghai 200232, China}

\author{Shoudong Shen}
\affiliation{State Key Laboratory of Surface Physics and Department of Physics, Fudan University, Shanghai 200433, China}

\author{Russell A. Ewings}
\affiliation{ISIS Pulsed Neutron and Muon Source, STFC Rutherford Appleton Laboratory, Harwell Campus, Didcot, Oxon, OX11 0QX, United Kingdom}

\author{Yang Zhao}
\affiliation{NIST Center for Neutron Research, National Institute of Standards and Technology, Gaithersburg, MD 20899, USA}
\affiliation{Department of Materials Science and Engineering, University of Maryland, College Park, MD 20742, USA}

\author{Leland W. Harriger}
\author{Jeffrey W. Lynn}
\affiliation{NIST Center for Neutron Research, National Institute of Standards and Technology, Gaithersburg, MD 20899, USA}

\author{Jun Zhao}\email[]{zhaoj@fudan.edu.cn}
\affiliation{State Key Laboratory of Surface Physics and Department of Physics, Fudan University, Shanghai 200433, China}
\affiliation{Institute of Nanoelectronics and Quantum Computing, Fudan University, Shanghai 200433, China}
\affiliation{Shanghai Qizhi Institute, Shanghai 200232, China}
\affiliation{Shanghai Research Center for Quantum Sciences, Shanghai 201315, China}

\pacs{25.40.Fq; 74.70.-b; 75.30.Ds; 75.30.Et}

\date{\today}

\begin{abstract}
The discovery of pressure-induced superconductivity in helimagnets (CrAs, MnP) has attracted considerable interest in understanding the relationship between complex magnetism and unconventional superconductivity. However, the nature of the magnetism and magnetic interactions that drive the unusual double-helical magnetic order in these materials remains unclear. Here, we report neutron scattering measurements of magnetic excitations in CrAs single crystals at ambient pressure. Our experiments reveal well defined spin wave excitations up to about 150~meV with a pseudogap below 7~meV, which can be effectively described by the Heisenberg model with \NN{} exchange interactions. Most surprisingly, the spin excitations are largely quenched above the N\'eel temperature, in contrast to cuprates and iron pnictides where the spectral weight is mostly preserved in the paramagnetic state. Our results suggest that the helimagnetic order is driven by strongly frustrated exchange interactions, and that CrAs is at the verge of itinerant and correlation-induced localized states, which is therefore highly pressure-tunable and favorable for superconductivity.
\end{abstract}

\maketitle

Significant efforts have been made to elucidate the nature of magnetism that is closely related to high-temperature superconductivity since the discovery of cuprates and \FeSC{}. The parent compounds of cuprates are Mott insulators with the N\'eel-type magnetic order \cite{Lee2006review}, and those of iron pnictides are semimetals with the stripe-type spin structure \cite{Dai2015Antiferro}. The recent discovery of Cr- and Mn-based superconductivity is surprising, given that the strong magnetism associated with Cr and Mn was previously thought to be unfavorable for superconductivity \cite{Wu2014Super,Cheng2015Pressure}. Both CrAs and MnP show incommensurate noncollinear helimagnetic order, unlike cuprate and \FeSC{} with simple collinear magnetic order. Superconductivity appears when external pressure that gradually suppresses the helical magnetic order is applied, implying a spin-fluctuation-mediated unconventional pairing mechanism \cite{Wu2014Super,Kotegawa2014Super,Cheng2015Pressure,Norman2015Super,Cheng2017Pressure,Chen2018Progress}. This is further supported by spin fluctuations and lack of coherence effect observed in the $^{75}$As-nuclear quadrupole resonance measurements on pressurized CrAs \cite{Kotegawa2015Detection}. However, the nature of the double-helical magnetic order remains unclear. Noncollinear or incommensurate magnetic order could originate from \gls*{DM} interactions \cite{Fert2013Skyrmion}, frustrated exchange couplings among localized spins \cite{Kallel1974Heli}, or spin-density-wave-like order of itinerant electrons \cite{Fawcett1988Spin}. Previous neutron scattering measurements on polycrystalline samples have shown spin excitations in CrAs \cite{Matsuda2018Evolution}. However, the detailed dispersion of the spin excitations and the associated magnetic interactions could not be unambiguously determined due to the limited information that can be obtained on the powder sample. How the seemly strong magnetism associated with Cr and Mn becomes compatible with superconductivity and what is the extent of electron correlation involvement in the magnetism are also unclear.

\begin{figure}
\includegraphics{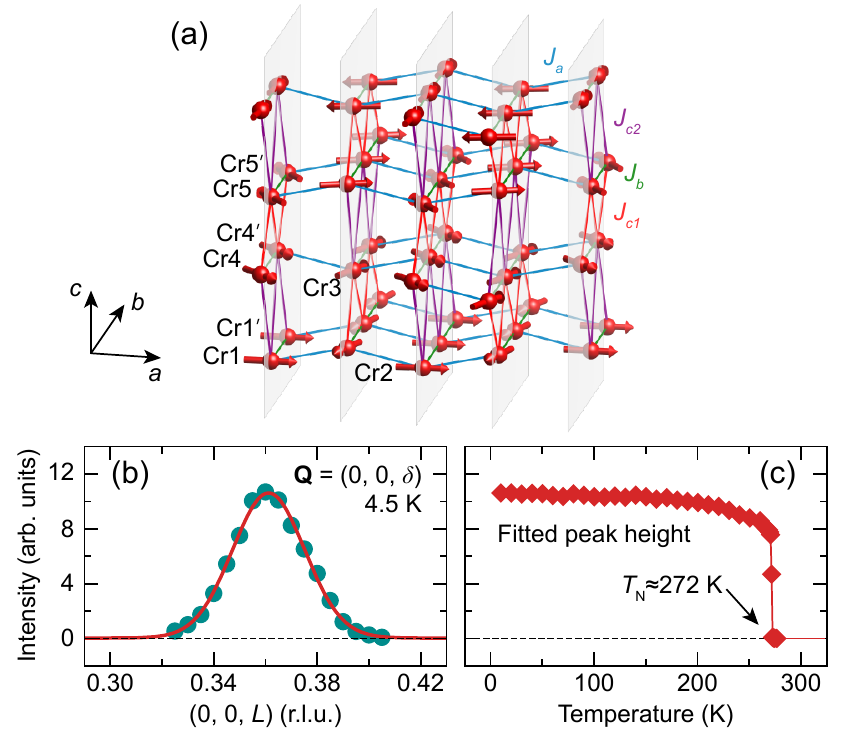}
\caption{Double helical magnetic order in CrAs. (a) Magnetic structure of CrAs at 5~K with spins rotating in the $ab$ plane. The four inequivalent \NN{} magnetic interactions included in our Heisenberg model are demonstrated as color-coded bonds. The arsenic ions are not shown. (b) Elastic \Q{} scan across the magnetic Bragg peak fitted with a pseudo-Voigt profile (solid line). (c) Temperature dependence of the fitted peak height of $\mathbf{Q}=(0, 0, \delta)$. The solid line is of guide for the eye.}
\label{fig:struct}
\end{figure}

To address these issues, we use inelastic neutron scattering to study the spin excitations in CrAs at ambient pressure. Our CrAs single crystals were grown using the flux method, as described elsewhere \cite{Wu2010Low}. It crystallizes to form an orthorhombic structure with the space group $Pnma$ (No.~62) and lattice constants of $a=5.6023$~\AA{}, $b=3.5866$~\AA{}, and $c=6.1322$~\AA{}. The wavevector is defined as $\mathbf{Q}=H2\pi/a+K2\pi/b+L2\pi/c$ in reciprocal lattice units (r.l.u.). The neutron scattering experiments were conducted on the BT-7 and SPINS triple-axis spectrometers at the NIST Center for Neutron Research \cite{Lynn2012Double}, and MAPS time-of-flight chopper spectrometer at the ISIS spallation neutron source \cite{Ewings2019Upgrade}. We fixed the final neutron energy at $E_{\mathrm{f}}=14.7$~meV and 5~meV for the BT-7 and SPINS measurements, respectively. More than 300 pieces of single crystals were co-aligned in the $(H, K, 0)$ horizontal scattering plane with total mass of $\sim4.5$ g and mosaicity of $\sim1^{\circ}$ for the triple-axis spectrometer measurements. For the MAPS experiment, $\sim7.5$ g of samples were measured with two incident energies, $E_{\mathrm{i}}=240$~meV and 500~meV, resulting in energy resolutions of 17.1~meV and 36.5~meV, respectively. The MAPS data were collected by rotating the samples and analyzed using the HORACE program \cite{HORACE}.

\textit{Results and discussion.} We start by characterizing the magnetic order in CrAs. Previous measurements have shown that CrAs exhibits a noncollinear double-helical magnetic phase transition, accompanied by a magnetostriction with a large expansion along the $b$ axis below $T_\mathrm{N}\approx 270 $ K \cite{Watanabe1969Magnetic,Selte1971Magnetic,Kazama1971Study,Boller1971First,Shen2016Structural,Keller2015Pressure,Pan2020Anomalous}. The Cr spin moments lie and rotate in the $ab$ plane and propagate along the structural $c$ axis [Fig.~\ref{fig:struct}(a)], resulting in satellite magnetic peaks at $\mathbf{Q}=\mathbf{G}\pm\mathbf{k}$, where $\mathbf{G}$ is the structural Bragg peaks and $\mathbf{k}=(0, 0, \delta)$ is the propagation vector. Our neutron diffraction measurements on single crystalline CrAs reveal $\delta\approx0.361$ at 4.5~K [Fig.~\ref{fig:struct}(b)], which is broadly consistent with the results reported in powder samples \cite{Shen2016Structural}. As the temperature increases, $\delta$ increases gradually and the peak intensity vanishes abruptly at $T_{\mathrm{N}}\approx272$~K [Fig.~\ref{fig:struct}(c)], consistent with previous reports \cite{Shen2016Structural}.

\begin{figure}
\includegraphics{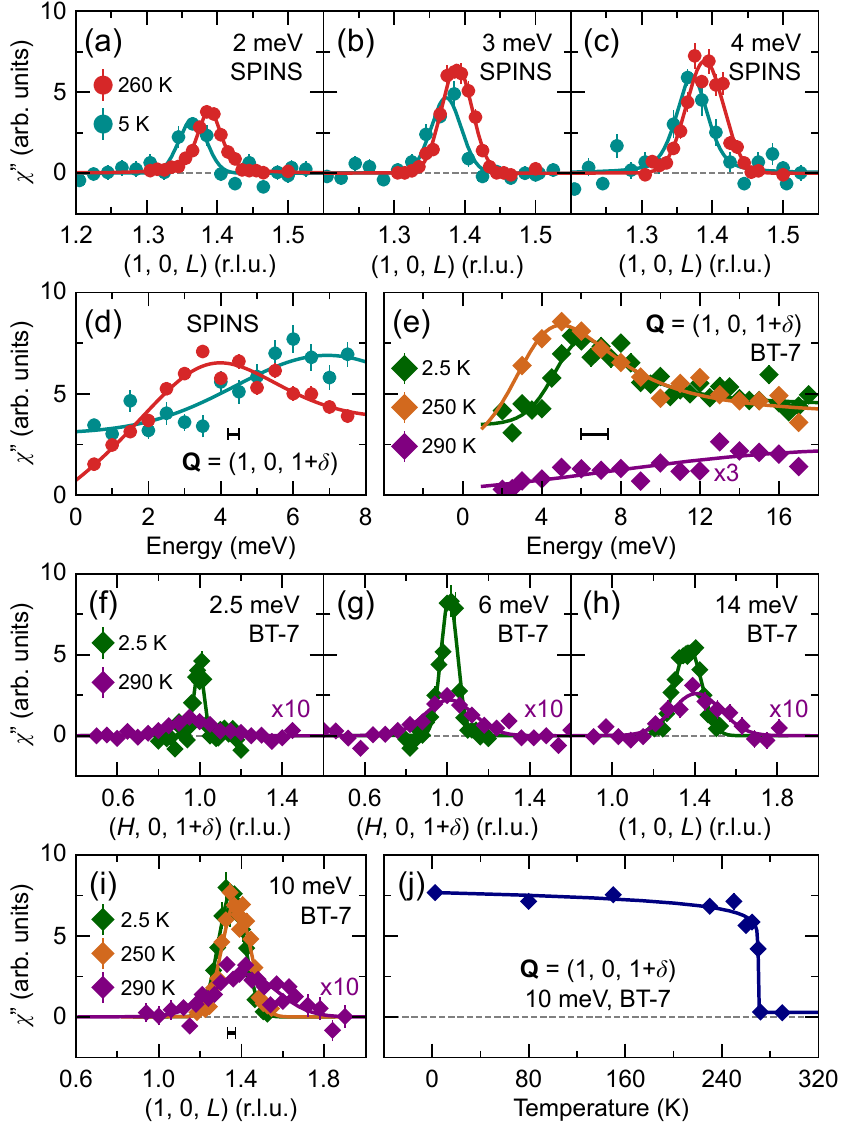}
\caption{Temperature dependence of the imaginary part of local dynamic susceptibility $\chi''$ in CrAs. (a)--(c) Bose-factor corrected constant energy scans across the magnetic peak $\mathbf{Q}=(1, 0, 1+\delta)$ at different energies. The data were fitted with pseudo-Voigt profiles (solid lines), and sloped backgrounds were subsequently subtracted. [(d),\ (e)] Background subtracted energy scans at $\mathbf{Q}=(1, 0, 1+\delta)$ with different temperatures. The backgrounds were measured at $\mathbf{Q}=(1, 0, 1.5)$ and $(1.5, 0, 1)$ for the SPINS and BT-7 data, respectively. The solid lines are of guides for the eye and the horizontal bars indicate the energy resolutions. Note that the intensities of 290~K data are multiplied by 3. (f)--(i) Bose-factor corrected \Q{} scans at different energies. Solid lines are the fitting results. The intensities of 290~K data are multiplied by 10. Error bars represent 1 standard deviation based on Poisson statistics. The horizontal bar in (i) represents the \Q{} resolution. (j) Temperature dependence of the fitted peak height of $\mathbf{Q}=(1, 0, 1+\delta)$ at 10~meV. The solid line is of guide for the eye.}
\label{fig:Tdep}
\end{figure}

\begin{figure*}
\includegraphics{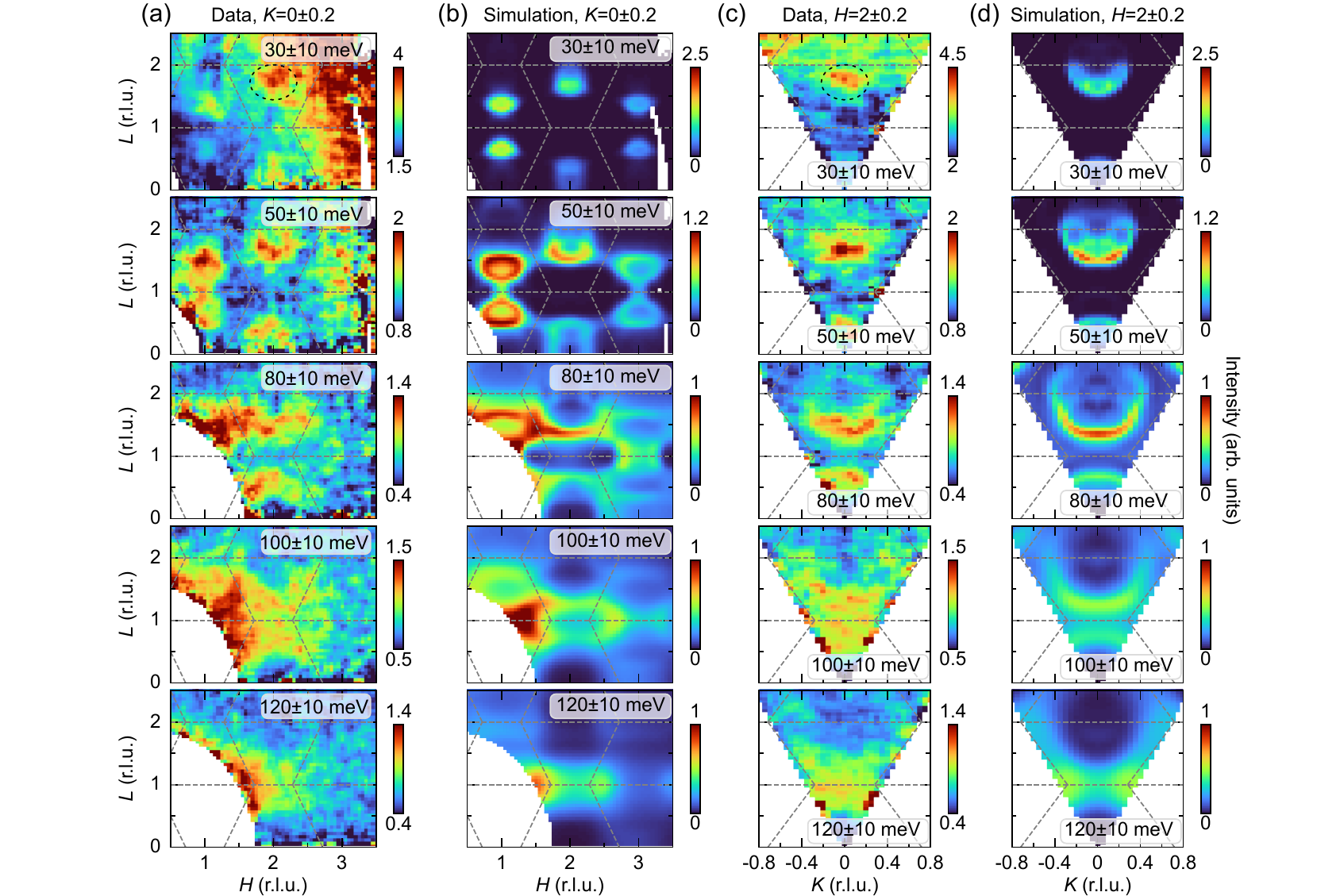}
\caption{Momentum dependence of the magnetic excitations in CrAs at 8~K. [(a),\ (c)] Constant-energy slices of the neutron-scattering intensities in $(H, 0, L)$ and $(2, K, L)$ planes at the indicated energies measured on the MAPS spectrometer. The data were symmetrized to enhance statistics and smoothed using a Savitzky--Golay filter. The strong intensities at high \Q{} originate from phonons of aluminum sample holders, and the white regions are momentum points that are not covered by neutron detectors. The 30, 50, and 80~meV data were collected with $E_{\mathrm{i}}=240$~meV, whereas the 100 and 120~meV data were obtained with $E_{\mathrm{i}}=500$~meV. The dashed ellipses mark the position of $\mathbf{Q}=(2, 0, 2-\delta)$. [(b),\ (d)] Calculated spin excitations using the model described in the text. The dashed lines indicate the magnetic zone boundaries.}
\label{fig:constE}
\end{figure*}

We move on to the inelastic neutron scattering measurements after confirming the magnetic order. We first use triple-axis spectrometers to measure the low energy spin excitations of CrAs. Sharp spin excitations are observed near the magnetic wavevector $\mathbf{Q}=(1, 0, 1+\delta)$ at 5~K [Figs.~\ref{fig:Tdep}(a)--\ref{fig:Tdep}(c)]. The energy-dependent spin-excitation spectrum shows a gap behavior below about 7~meV, but the excitations persist down to the lowest measured energy, exhibiting a pseudogap feature [Figs.~\ref{fig:Tdep}(d) and \ref{fig:Tdep}(e)]. This is different from the parent compounds of cuprate and iron pnictide superconductors with clean spin gaps. The pseudogap behavior could be due to the presence of a weak easy-plane anisotropy that constrains the spins to the $ab$ plane, and the finite in-gap spectral weights originate from the excitations within the plane.

\begin{figure*}
\includegraphics{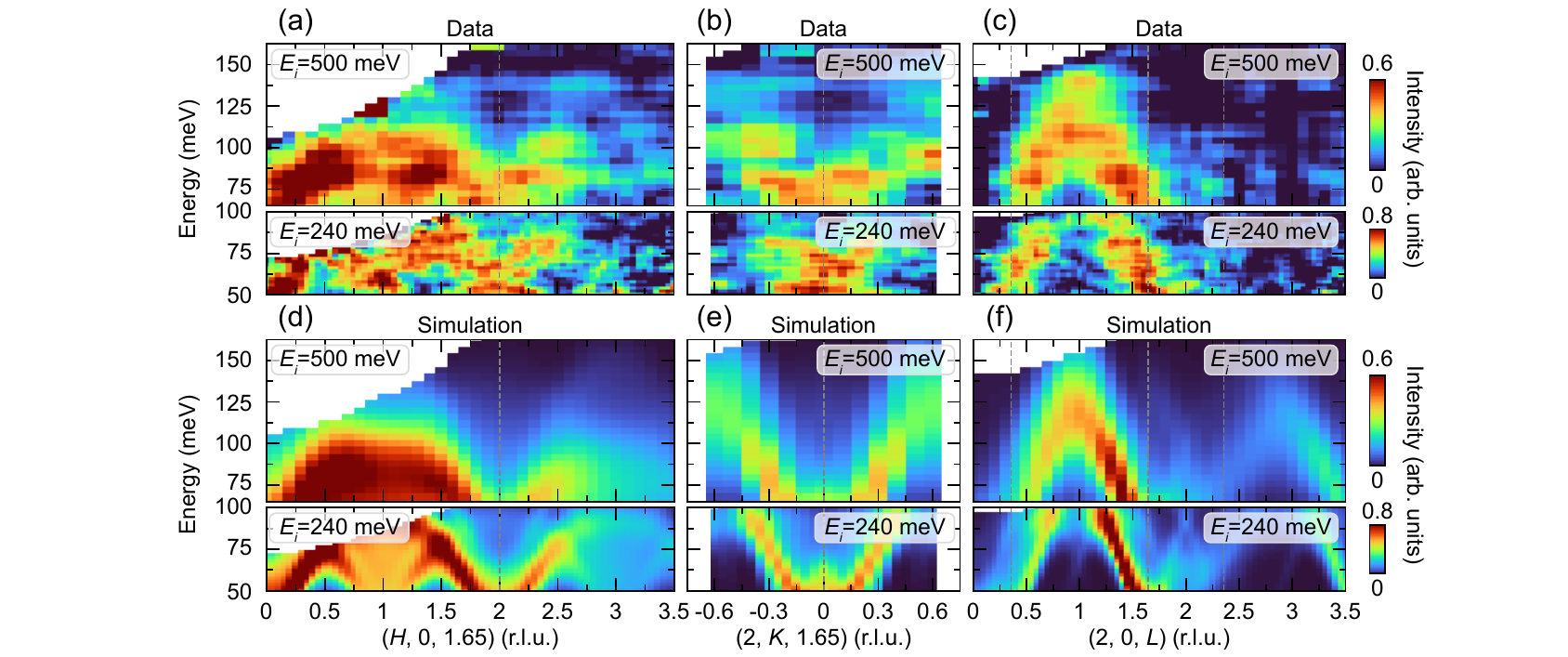}
\caption{Energy dependence of the magnetic excitations in CrAs along high-symmetry momentum directions at 8~K. (a)--(c) \textit{E}--\Q{} plots of the neutron scattering intensities across $\mathbf{Q}=(2, 0, 2-\delta)$. An energy-dependent background was subtracted using the scattering near $\mathbf{Q}=(2, 0, 3)$ as a reference, which may result in over-subtraction in some areas. Only the data above the phonon cutoff energy are presented. (d)--(f) Simulated spin-wave dispersions using the model described in the text. The vertical dashed lines indicate the magnetic zone centers.}
\label{fig:disp}
\end{figure*}

Figure~\ref{fig:constE} shows the momentum dependence of the high-energy spin excitations of CrAs measured on the MAPS time-of-flight spectrometer. Pronounced spin excitations originating from the magnetic zone centers disperse outward with increasing energy and merge near the zone boundaries. Interestingly, Fig.~\ref{fig:constE}(c) shows that the intensity distribution of the spin waves exhibits spatial anisotropy along the $L$ direction, with significantly stronger intensity near $L=1$. This can be observed more clearly in the \textit{E}--\Q{} plots along high symmetry directions in Figs.~\ref{fig:disp}(a)--\ref{fig:disp}(c). The magnetic excitations remain coherent up to the band top of $\sim$150~meV. 

We attempt to implement the Heisenberg model to quantitatively determine the magnetic interactions of CrAs:
\begin{equation}
	\mathcal{H} = \sum_{\langle ij \rangle}{J_{ij} \mathbf{S}_i \cdot \mathbf{S}_j}
\end{equation}
where $J_{ij}$ includes four \NN{} exchange interactions along different directions, $J_{a}$, $J_{b}$, $J_{c1}$, and $J_{c2}$, and $\langle ij \rangle$ denotes the corresponding bonds, as illustrated in Fig.~\ref{fig:struct}(a). Two inequivalent interactions, $J_{c1}$ and $J_{c2}$, run alternately along the $c$ axis. The SPINW program is used to analyze the spin excitation spectrum and magnetic structure \cite{SPINW}. Figures~\ref{fig:constE}(b),\ \ref{fig:constE}(d), and \ref{fig:disp}(d)--\ref{fig:disp}(f) depict the calculated neutron spectra convoluted with energy resolution using the optimized parameters in Table~\ref{table:Js}, which show great consistency with the data, including the anisotropic intensity distribution along the $L$ direction. The corresponding magnetic structure to this set of exchange parameters can be solved analytically and has a propagation vector of $\mathbf{k}=(0, 0, 0.36)$, in good agreement with the diffraction results, putting it in a strong frustration regime of the phase diagram \cite{Kallel1974Heli}. The calculated phase angle of $\sim-126^{\circ}$ between the two helical chains is also close to that determined by the neutron diffraction experiment ($\phi_{12}\sim-109^{\circ}$) \cite{Shen2016Structural}. These results suggest that the Heisenberg model can effectively explain the detailed magnetic structure and the associated spin wave excitations in CrAs.

More insights into the nature of the magnetic interactions can be obtained by analyzing the Cr-Cr bond lengths and the associated exchange interactions \cite{Khomskii2014Transition}. We note that $J_b$, $J_{c1}$, and $J_{c2}$ bonds involve two edge-sharing CrAs$_6$ octahedra. For Cr$^{3+}$ with $t^3_{\mathrm{2g}}$ configuration, the superexchange interaction induced by the virtual electronic hopping through the ligand As is \gls*{FM}, whereas the exchange interaction through direct Cr-Cr hopping is \gls*{AFM}. The competition between these two contributions varies sensitively depending on the Cr-Cr bond length. For a short Cr-Cr bond, the total interaction $J$ is strongly \gls*{AFM} since the direct exchange dominates. $J$ decreases as the Cr-Cr bond length increases, and eventually turns into \gls*{FM} when the direct exchange is overtaken by the superexchange interaction. This analysis is consistent with our simulated exchange couplings based on the spin wave excitation spectrum in CrAs (Table~\ref{table:Js}). For instance, $J_{c1}$, with the shortest Cr-Cr bond ($\sim$3.09~\AA{}) is strongly \gls*{AFM}, $J_b$ ($\sim$3.5883~\AA{}) is weakly \gls*{AFM}, and $J_{c2}$ ($\sim$4.042~\AA{}) is weakly \gls*{FM}. A similar situation also applies to $J_a$, involving two face-sharing CrAs$_6$ octahedra. Despite the weak \gls*{AFM} $J_b$, the spin moments of Cr4 and Cr4' are ferromagnetically aligned in CrAs [Fig.~\ref{fig:struct}(a)]. This is because Cr4 and Cr4' are also connected to Cr1 and Cr5 through $J_{c2}$ and $J_{c1}$, respectively, which enforces the parallel alignment.

It has been suggested that the \gls*{DM} interactions may also drive a helimagnetic order in binary pnictides \cite{Cuono2022Double}. We calculate the symmetry-allowed \gls*{DM} interaction unit vector $\mathbf{D}$ for all four \NN{} bonds involved in our model in CrAs \cite{Keffer1962Moriya}, which gives $\mathbf{D}_a=(0, 1, 0)$, $\mathbf{D}_b\approx(0.73, 0, 0.68)$, and $\mathbf{D}_{c1}=\mathbf{D}_{c2}=0$. Because the Cr spin moments lie in the $ab$ plane and are ferromagnetically aligned along the $b$ direction, the contributions of \gls*{DM} interactions to the Hamiltonian, $\mathbf{D}_{ij} \cdot (\mathbf{S}_i \times \mathbf{S}_j)$, vanish for all four terms in the ground state. This is consistent with our analysis, suggesting that a localized Heisenberg model with \NN{} exchange interactions is sufficient to explain spin wave excitations at ambient pressure. Our previous neutron diffraction experiments have revealed a spin reorientation from the $ab$ plane to the $ac$ plane under an external pressure of $P=0.6$~GPa \cite{Shen2016Structural}. It is likely that the contribution from $\mathbf{D}_a$ at the critical pressure is strong enough to overcome the small easy-$ab$-plane anisotropy and to make the $ac$ plane more favorable.

Static magnetic order usually competes with superconductivity but the fluctuations related to the magnetic order could mediate the electron pairing. The paramagnetic spin excitations of the parent compounds of unconventional superconductors are often similar to those of their superconducting counterparts with suppressed magnetic order \cite{Dai2015Antiferro,Harriger2011Nematic}, which could have substantial implications for the superconductivity mechanism. In CrAs at ambient pressure, the pseudo spin gap softens as temperature increases [Figs.~\ref{fig:Tdep}(d) and \ref{fig:Tdep}(e)]. Most surprisingly, only very weak short-range correlated incommensurate spin fluctuations survive above \TN{} near the similar magnetic wavevector as in the ordered state [Figs.~\ref{fig:Tdep}(e)--\ref{fig:Tdep}(i)]. The detailed temperature dependence measurements show that the vast majority of the spectral weight is abruptly quenched at \TN{} [Fig.~\ref{fig:Tdep}(j)], distinct from cuprates and iron pnictides where the spectral weight is largely preserved in the paramagnetic state \cite{Harriger2011Nematic,Zhao2014Neutron,Dai2015Antiferro,Sapkota2021Reinvestation}. 

Theoretical calculations suggest that CrAs is a weakly correlated metal with Coulomb repulsion of 0.37--1~eV \cite{Autieri2017First,Autieri2018Tight,Cuono2022Double}, which is much lower than those of iron pnictides and cuprates. Therefore, it is quite intriguing that a localized Heisenberg model can accurately explain the detailed spin excitation spectrum and helimagnetic structure of CrAs below \TN{}. The orbital-selective Mottness \cite{Yu2011Mott,Medici2014Selective} proposed to influence \FeSC{} may also play a role here. It is possible that some Cr $t_{\mathrm{2g}}$ orbitals contribute to the metallic transport behavior, whereas others remain largely localized and render exchange interactions responsible for the helimagnetism in CrAs. On warming across \TN{}, accompanied by the release of magnetostriction, the localized electrons may become more itinerant due to the increase of band width on the associated orbitals (such as $d_{yz}$) caused by the significant lattice shrinkage along the $b$ axis \cite{Ito2007Electronic}. Consequently, the local moments are largely quenched and the spin excitation intensity is significantly suppressed above \TN{}. These results suggest that CrAs is at the verge of itinerant and correlation-induced localized states, which together with the frustrated magnetic interactions makes the magnetism of CrAs highly pressure-tunable. Thus, although the magnetism in Cr-based compounds is often too strong and detrimental to superconductivity, the helimagnetic order in CrAs is actually quite fragile and can be suppressed by moderated pressure, which is favorable for superconductivity. The significantly suppressed spin fluctuations in the paramagnetic state could also explain the relatively low \Tc{} observed in CrAs, given the comparable magnetic interaction energy scales among iron pnictides, cuprates and CrAs.

\begin{table}
\caption{Magnetic exchange coupling constants determined for CrAs at 8~K and ambient pressure.}
\centering
\begin{ruledtabular}
\begin{tabular}{cccc}
$SJ_a$ (meV) & $SJ_b$ (meV) & $SJ_{c1}$ (meV) & $SJ_{c2}$ (meV) \\ 
\hline
15(2) & 5(4) & 34(5) & -7(1) \\
\end{tabular}
\end{ruledtabular}
\label{table:Js}
\end{table}

In summary, we have investigated spin excitations in helimagnet CrAs using inelastic neutron scattering. In the magnetically ordered state, the observed strong spin excitations remain coherent up to the band top of 150~meV.  The spin waves can be effectively described using the Heisenberg model with four \NN{} magnetic exchange interactions, and the competition between them results in the double-helical magnetic structure. In contrast to the clean spin gap commonly observed in cuprates and iron pnictides, CrAs shows a pseudogap behavior below 7~meV, exhibiting unique spin excitations associated with the helical order and easy-plane anisotropy. On warming to above \TN{}, the spin fluctuations are largely quenched and only very weak short-range correlated incommensurate antiferromagnetic spin fluctuations survive. Our results suggest that CrAs is at the verge of itinerant and correlation-induced localized states with a fragile helimagnetic order, which therefore can be easily tuned by moderate pressure and is favorable for superconductivity.

This work was supported by the National Natural Science Foundation of China (Grant No.~11874119), and Shanghai Municipal Science and Technology Major Project (Grant No.~2019SHZDZX01). The datasets for the inelastic neutron scattering experiment on the MAPS chopper spectrometer are available from the ISIS facility, Rutherford Appleton Laboratory data portal (10.5286/ISIS.E.RB1510119). The neutron data taken at the NCNR are available at ftp://ftp.ncnr.nist.gov/pub/ncnrdata. All other data that support the plots within this work and other findings in this study are available from the corresponding authors upon reasonable request.









\bibliography{refs}

\end{document}